\documentclass[a4paper,twocolumn,floatfix,showpacs,showkeys,
groupedaddress]{revtex4}

\usepackage{bm,amsmath,amssymb,amsfonts}

\usepackage[dvips]{graphicx}

\usepackage{color}
\definecolor{blue}{rgb}{0.0,0.0,0.6}
\begin{document}

\title{\textcolor{blue}
{Absolutely continuous energy bands in the 
electronic spectrum of \\quasiperiodic ladder networks}}

\author{Biplab Pal}
\email{biplabpal@klyuniv.ac.in}

\author{Arunava Chakrabarti}
\email{arunava\_chakrabarti@yahoo.co.in}
\thanks{Tel.: +91 33 25820184, Fax: +91 33 25828282}

\affiliation{Department of Physics, University of Kalyani, Kalyani,
West Bengal-741 235, India}

\begin{abstract}
The energy spectra of quasi-one dimensional quasiperiodic ladder networks are 
analyzed within a tight binding description. In particular, we show that if a 
selected set of sites in each strand of a ladder is tunnel-coupled to quantum 
dots attached from a side, absolutely continuous subbands can be generated in 
the spectrum if one tunes the dot potential and the dot-strand coupling appropriately. 
Typical  cases with two and three strand Fibonacci ladders in the off diagonal 
model are discussed in details. We also discuss the possibility of re-entrant 
insulator-metal transition for a general $n$-strand ladder network when $n$ becomes 
large. The observations remain valid even in the case of a disordered ladder network 
with the same constituents. The results are analytically exact. 
\end{abstract}
\maketitle
%%%%%%%%%%%%%%%%%%%%%%%%%%%%%%%%%%%%%%%%%%%%%%%%%%%%%%%%%%%%%%%%%%%%%%%%%%%%%%%%%%
\section{Introduction}
\label{sec1}
Anderson localization~\cite{anderson,kramer,abrahams} of electronic states 
in a disordered system is a path breaking observation that has extended 
its realm well beyond the electronic properties of disordered solid 
materials, and has been found out to be ubiquitous in a wide variety of 
systems such as the photonic~\cite{yablo,john}, 
phononic~\cite{montero,vasseur}, polaronic~\cite{barinov,grochol}, 
or plasmonic lattices~\cite{tao,christ} to name a few. Artificial, 
tailor made geometries developed using the improved fabrication 
and lithographic techniques have been instrumental in testing the 
validity of the basic theory of exciton localization in presence of 
disorder. Examples include the direct observation of localization of 
matter waves in recent times~\cite{damski,billy,roati}.

Localization is a result of quantum interference that has the 
strongest manifestation in one dimension where all the single 
particle states exhibit an exponential decay of amplitude as one 
moves away from a particular lattice point. For dimensions beyond 
one, the states retain their characteristic decay in amplitude, 
with the addition of a possibility of a metal-insulator transition 
in three dimensions. The results have been substantiated by 
calculations of the localization length~\cite{rudo1,rudo2}, 
density of states~\cite{alberto1} and their multi-fractality~\cite{rudo3,rudo4}. 

Extensive work has also revealed the intricacies of the single 
parameter scaling hypothesis~\cite{abrahams}, its validity~\cite{rudo5} 
or, variance~\cite{deych}, and even violation~\cite{titov,bunde,kravtsov} in low 
dimensional systems within a tight binding approximation. The studies 
have subsequently been consolidated by direct experimental measurements 
of conductance distribution in quasi-one dimensional gold wires~\cite{mohanty}.

In the last couple of decades variations in the canonical cases of 
Anderson localization have surfaced, particularly, within a tight 
binding description. Resonant tunnelling of electronic states has 
been shown to arise even in a one dimensional disordered system as 
a consequence of certain special kinds of positional 
correlation~\cite{dunlap,moura,maiti,rudo6}. Discrete energy levels 
corresponding to {\it extended eigenfunctions}~\cite{dunlap}, appear 
within a typical point spectrum exhibited by the randomly disordered 
lattices rendering the two-terminal transport completely ballistic at 
special values of the Fermi level. Such results are common even to 
infinite quasicrystalline lattices in one dimension where, the self similar 
character of the chain really allows an infinite number of unscattered (extended) 
eigenfunctions despite the absence of any translational invariance~\cite{arunava1}.

In this context, rather unusual, yet meaningful question could be, 
is it possible to engineer, in 
a controlled fashion, an {\it absolutely continuous} portion in the energy spectrum 
of non-translationally invariant systems with the help of some kind positional 
correlations ?
The  
existence of real continuous bands are reported recently in 
quasi one dimensional or two dimensional systems only with diagonal 
disorder~\cite{maiti,rudo6}. We thus have a partial answer to the question, 
and it remains to be seen whether such 
bands of extended eigenfunctions can be generated and controlled with 
off-diagonal disorder are well.

We address this issue with respect to quasiperiodically ordered geometries. 
In this communication we show that, in quasi-one dimensional ladder networks 
with off-diagonal quasi-periodic order it is possible to engineer absolutely 
continuous bands of energy eigenvalues. We specifically work with the 
Fibonacci quasiperiodic order, though the observation is by no means, 
restricted to them. Ladder networks within the framework of tight binding 
model have already been exploited to gain insight into charge transport in 
DNA~\cite{macia}, to unravel interesting physics related to polymer structures 
and ployacene lattice topology~\cite{grosso1,grosso2}. Comparatively 
speaking, controlled generation of  absolutely continuous portions in 
the energy spectrum of the ladder networks, disordered or quasi-periodic, 
still remains to be explored. If it works then such an effort may open up 
new physics, or allow one to explore the possibility of devising novel 
low dimensional filter networks.

The central result of this paper is that, one can {\it create} absolutely 
continuous subbands in the electronic spectrum of such quasi 1-d systems, 
by attaching a single quantum dot (QD) from one side to a selected set of 
lattice points in every strand of the ladder. We discuss the results explicitly 
in terms of Fibonacci quasi-periodic two and three arm ladder network models 
using a tight binding Hamiltonian and a real space renormalization group 
(RSRG) method. We also investigate whether a crossover in the fundamental 
character of the eigenfunctions, going from a typical {\it critical} nature 
to a Bloch-like {\it extended} character can be engineered. As will be 
shown, this indeed turns out to be quite a possibility in the case of the 
quasi-periodic ladder network (QPLN), particularly when the number of 
strands `$n$' in the transverse direction increases. The results hold even 
for an off-diagonally disordered ladder where a re-entrant 
insulator-metal-insulator transition becomes feasible.

The system is described by a tight binding Hamiltonian, and is depicted in 
Fig.~\ref{lattice1}. The on-site potential at every lattice point is kept 
constant, while the nearest neighbor hopping integral along the arms of the 
ladder assumes two values dictated by the growth rule of the desired 
quasi-periodic sequence. The electron can tunnel between the arms of the 
ladder by hopping from one site to its corresponding site on the other arm 
through the hopping integral $\xi$ (see Fig.~\ref{lattice1}). The sites 
marked by $\alpha$ in each arm of the ladder is coupled to a single level 
QD~\cite{kubala} attached from one side. The QD couples to every 
$\alpha$-site via a hopping integral $\lambda$. There is no positional 
correlation, short range or long range in the conventional sense~\cite{dunlap,moura}. 
Attachment of QD's from a side can quench the conductance of a 
linear chain~\cite{grosso3}, but an array of QD's in a ladder can have 
reverse effect, as will be seen.

We will show in a completely analytical way that, for a particular relationship 
between the nearest neighbor hopping integrals along the ladder strand and the 
QD-strand coupling $\lambda$, and for a special value of the QD potential, we 
get {\it absolutely continuous subbands} in the spectrum. This immediately 
raises a non-trivial issue that, the nature of the electronic spectrum and 
the eigenstates over a {\it continuous range of eigenvalues} can become 
sensitive to the {\it numerical values} of the Hamiltonian parameters -- a fact 
that is not commonly observed in the conventional Anderson localization problem. 
The full spectrum turns out to be an admixture of conventional {\it critical 
eigenstates}, characteristics of a Fibonacci quasiperiodic chain, and absolutely 
continuous parts, giving rise to the possibility of a {\it phase transition}, 
particularly when the system grows in the transverse direction.  
Our results are analytically exact.

In section~\ref{sec2} we describe the Hamiltonian and the basic working 
principle for the two and three strand quasi-periodic ladders constructed 
following a Fibonacci sequence. Along with this, the real space renormalization 
scheme to obtain the energy spectrum and to judge the character of the single 
particle states is also presented. In 
section~\ref{sec3} we present a result for the two terminal 
transmission coefficient of a finite sized QPLN, and 
in section~\ref{sec4} we draw our conclusion.
%%%%%%%%%%%%%%%%%%%%%%%%%%%%%%%%%%%%%%%%%%%%%%%%%%%%%%%%%%%%%%%%%%%%%%%%%%%%%%%%%%
\section{The Model and the method}
\label{sec2}
In Wannier
basis the tight binding Hamiltonian of a general $n$-arm 
ladder network reads~\cite{rudo6},
\begin{equation}
\mbox{\boldmath $H$}  = \sum_{i} \mbox{\boldmath $\epsilon_{i} c_{i}^{\dagger} c_{i}$}
+\sum_{\langle ij \rangle} \mbox{\boldmath $t_{ij}$} \left[\mbox{\boldmath $c_{i}^{\dagger} 
c_{j}$} 
+ h.c. \right]
\end{equation}
\label{hamilton}
where, $\mbox{\boldmath $\epsilon_{i}$}$ is, in general, an $n \times n$ on-site 
potential matrix at the $i$-th vertical rung. We shall explicitly discuss the two-strand 
ladder at first.    
The cases $n \ge 2$ can be trivially constructed.
$\mbox{\boldmath $c_{i}^\dag (c_{i})$}$ is the creation (annihilation) 
operator, represented by rows (columns) of appropriate dimension, 
and $\mbox{\boldmath $t_{ij}$}$ is the $n \times n$ hopping matrix representing 
hopping along and in between strands of the ladder, $n$ being equal to $2$, 
$3$ or any integer.

The Fibonacci ladder network is grown by attaching identical one dimensional 
Fibonacci chains in the transverse direction. Each individual chain 
grows along the horizontal direction following the binary Fibonacci 
sequence of two letters $L$ and $S$ (representing two bonds). 
The growth rule is, $L \rightarrow LS$ and 
$S \rightarrow L$, and the sequence begins with $L$. The on-site 
potentials at the vertices in each strand are designated by three symbols viz., 
$\epsilon_{\alpha}$, $\epsilon_{\beta}$ and $\epsilon_{\gamma}$ for 
sites flanked by two $L$-bonds, $L$ on left and $S$ on right, and 
$L$ on right and $S$ on left respectively. The attached QD is 
shown as ``$\mu$" in Fig.~\ref{lattice1}, and the potential at the side 
coupled QD is designated by $\epsilon_{\mu}$. The nearest neighbor 
hopping integral $t_{ij} = t_{L}$ or $t_{S}$ along ladder-arms following 
the Fibonacci sequence. $t_{ij} = \xi$ is the inter-arm hopping, 
and $t_{ij} = \lambda$ designates the tunnel hopping between the 
QD and the $\alpha$-sites of the ladder network. In our work, 
we shall assume 
$\epsilon_{\alpha}=\epsilon_{\beta}=\epsilon_{\gamma}=\epsilon_{0}$, and 
$t_{L} \ne t_{S}$. So, this is purely 
a {\it transfer model}~\cite{kohmoto} in the standard language of one dimensional
quasicrystals. 

We shall analyze the problem  using a method described earlier~\cite{maiti,rudo6} 
in which an `$n$'-strand ladder ($n=2$ in the first phase of our discussion) 
is decoupled, making a change of basis, into a system of $n$ independent 
linear chains. The complete spectrum of the quasi $1$-d system of $n$ strands 
is then obtained by convolving the spectra of the individual linear chains. 
In addition, we perform an RSRG analysis, which does not need any decoupling 
and yields a density of states spectrum that is in conformity with our results 
obtained using the decoupling scheme. The two terminal transmission spectrum 
is also computed using the RSRG decimation scheme to highlight the effect 
of the side coupled QD's.

The analysis will be carried over to the three strand ladder, and a 
generalization to the $n$ strand case will be discussed using arguments. 
%%%%%%%%%%%%%%%%%%%%%%%%%%%%%%%%%%%%%%%%%%%%
\subsection{The two-strand Fibonacci ladder}
\subsubsection{The decoupling of the strands and the spectral analysis}

At first, we renormalize the potential at the $\alpha$-sites by 
decimating the side coupled QD. This leads to a {\it renormalized} value of the 
on-site potential at the $\alpha$-vertex, given by, 
$\bar{\epsilon}_{\alpha} = \epsilon_{0} + \lambda^2/(E -\epsilon_{\mu})$.
%*************************************************
\begin{figure}[ht]
\centering
\includegraphics[clip,width=8cm,angle=0]{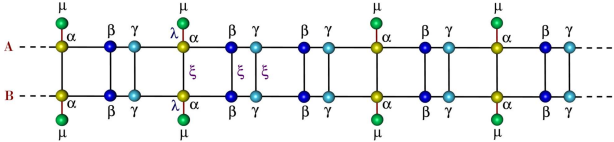}
\caption{(Color online) Typical realization of a two strand 
Fibonacci ladder. The sites $\alpha$, $\beta$ and $\gamma$ are colored 
differently to distinguish between the nearest neighbor bond-environments.
The QD's $\mu$ are coupled to the $\alpha$-sites in each case. The 
coupling between $\mu$ and $\alpha$-site is denoted by $\lambda$ and the 
inter-strand coupling is denoted by $\xi$.}
\label{lattice1}
\end{figure}
%*************************************************

The Schr\"{o}dinger equation for the two-arm ladder can now be  written 
equivalently in the form of the difference equations involving $2 \times 2$ 
matrices: 
\begin{equation}
( E \mbox{\boldmath $I$} - \mbox{\boldmath $\epsilon_{i}$} ) \mbox{\boldmath 
$\psi_{i}$} = \sum_{j} \mbox{\boldmath $t_{ij}$} \mbox{\boldmath $\psi_{j}$}
\label{difference}
\end{equation}
where, 
\begin{equation}
\mbox{\boldmath $\psi_{i}$} =   
\left ( \begin{array}{c}
\psi_{i,A} \\
\psi_{i,B}  
\end{array} \right )
\end{equation}
and, 
\begin{equation}
\mbox{\boldmath $\epsilon_{i}$} = 
\left( \begin{array}{cccc}
\epsilon_{i} & \xi \\ 
\xi & \epsilon_{i} 
\end{array}
\right)
\end{equation} 
and, 
$\mbox{\boldmath $I$}$ is the $2 \times 2$ unit matrix.
Written explicitly, the matrix equation reads, 
%%%%%%%%%%%%%%%%%%%%%%
\begin{widetext}
\begin{equation}
\left [
\left( \begin{array}{cccc}
E & 0 \\ 
0 & E 
\end{array}
\right)
- 
\left( \begin{array}{cccc}
\epsilon_{i} & \xi \\ 
\xi & \epsilon_{i} 
\end{array}
\right)
\right ]
\left ( \begin{array}{c}
\psi_{i,A} \\
\psi_{i,B}  
\end{array} \right )
= 
\left( \begin{array}{cccc}
t_{i,i+1}^A & 0 \\ 
0 & t_{i,i+1}^B
\end{array}
\right )
\left ( \begin{array}{c}
\psi_{i+1,A} \\
\psi_{i+1,B}  
\end{array} \right )
 +
\left( \begin{array}{cccc}
t_{i,i-1}^A & 0 \\ 
0 & t_{i,i-1}^B
\end{array}
\right )
\left ( \begin{array}{c}
\psi_{i-1,A} \\
\psi_{i-1,B}  
\end{array} \right )
\label{differenceeq}
\end{equation}
\end{widetext}
%%%%%%%%%%%%%%%%%%%%%%  
In Eq.~\eqref{differenceeq}, the potential $\epsilon_{i}$ takes on values 
$\bar{\epsilon}_{\alpha}$ corresponding to a 
renormalized $\alpha$-site in each arm, 
while it is equal to $\epsilon_{0}$ for both the $\beta$ and $\gamma$-sites 
in each arm. The nearest neighbor hopping integrals $t_{i,i\pm 1}^{A(B)}$ are 
$t_{L}$ or $t_{S}$ depending on the local environment of the corresponding site 
in the arm $A(B)$. A look at the Fig.~\ref{lattice1} will make it obvious that 
there are three such equations corresponding to the three different kinds of 
vertices viz., $\alpha$, $\beta$ and $\gamma$ respectively. The potential matrix 
$\mbox{\boldmath $\epsilon_{i}$}$ 

corresponding to the renormalized $\alpha$-sites is, 
\begin{equation}
\left( \begin{array}{cccc}
\epsilon_{0}+\dfrac{\lambda^{2}}{E-\epsilon_{\mu}} & \xi \\ 
\xi & \epsilon_{0}+\dfrac{\lambda^{2}}{E-\epsilon_{\mu}} 
\end{array}
\right)
\equiv
\mbox{\boldmath $\epsilon_{0}$}+
\dfrac{\lambda^{2}}{E-\epsilon_{\mu}} \mbox{\boldmath $I$}
\end{equation}
and that for the $\beta$ and $\gamma$ sites are 
\begin{equation}
\mbox{\boldmath $\epsilon_{0}$}
= \left( \begin{array}{cccc}
\epsilon_{0} & \xi \\ 
\xi & \epsilon_{0}
\end{array}
\right)
\end{equation}
We now diagonalize the $\mbox{\boldmath $\epsilon_{0}$}$
matrix using a matrix $\mbox{\boldmath $S$}$. The matrix $\mbox{\boldmath $S$}$ 
makes a change over to a new basis defined by,  
$\mbox{\boldmath $\phi_{i} = S^{-1} \psi_{i}$}$. 
As shown elsewhere~\cite{maiti,rudo6}, such a change of basis leads to a 
decoupling of the set of Eq.~\eqref{differenceeq} into two independent 
equations, one corresponding to the $A$-arm alone, and the other, to 
the arm $B$. Written explicitly, the equations read,   
\begin{eqnarray}
&\left [ E - \left(\epsilon_{0}+\xi+\dfrac{\lambda^2}{E-\epsilon_{\mu}}\right) 
\right ] \phi_{i,A}  = 
t_{L} \phi_{i+1,A} + t_{L} \phi_{i-1,A} \nonumber \\
&\left [E - (\epsilon_{0}+\xi) \right] \phi_{i,A}  =  
t_{S} \phi_{i+1,A} + t_{L} \phi_{i-1,A} \nonumber \\
&\left [E - (\epsilon_{0}+\xi) \right ] \phi_{i,A}  = 
t_{L} \phi_{i+1,A} + t_{S} \phi_{i-1,A} 
\label{arm1}
\end{eqnarray}
for the sites $\alpha$, $\beta$ and $\gamma$ sequentially in arm $A$. 
Similar set of equations for $\alpha$, $\beta$ and $\gamma$ sites in arm $B$ will be, 
\begin{eqnarray}
&\left [ E - \left(\epsilon_{0}-\xi+\dfrac{\lambda^2}{E-\epsilon_{\mu}}\right) 
\right ] \phi_{i,B}  =  
t_{L} \phi_{i+1,B} + t_{L} \phi_{i-1,B} \nonumber \\
&\left [E - (\epsilon_{0}-\xi) \right ] \phi_{i,B}  =  
t_{S} \phi_{i+1,B} + t_{L} \phi_{i-1,B} \nonumber \\
&\left [E - (\epsilon_{0}-\xi) \right ] \phi_{i,B}  =  
t_{L} \phi_{i+1,B} + t_{S} \phi_{i-1,B} 
\label{arm2}
\end{eqnarray}

Needless to say, each of the sets Eq.~\eqref{arm1} and Eq.~\eqref{arm2} 
separately represents an independent Fibonacci chain. The spectrum of each chain 
bears the usual singular continuous character. The spectrum of the original 
ladder network can be obtained by convolution of the density of states of 
the individual decoupled Fibonacci chains. However, an analysis of these 
decoupled sub-systems yields rich spectral insight that would otherwise be 
difficult to obtain. We are now in a position to throw some light on this issue.

First, we need to appreciate that, in the new basis each decoupled 
Fibonacci chain is described by a set of three {\it transfer matrices}, viz., 
$\mbox{\boldmath $M_{\alpha}$}$, $\mbox{\boldmath $M_{\beta}$}$ and 
$\mbox{\boldmath $M_{\gamma}$}$~\cite{kohmoto}. For the $A$-arm decoupled Fibonacci 
chain, the explicit expressions of these matrices are, 
\begin{eqnarray}
\mbox{\boldmath $M_{\alpha,A}$} &=& \left( \begin{array}{cccc}
[E - (\epsilon_{0}+\xi+\dfrac{\lambda^2}{E-\epsilon_{\mu}})]/t_{L} & -1 \\ 
1 & 0 
\end{array}
\right) \nonumber \\
\mbox{\boldmath $M_{\beta,A}$} &=& \left( \def\arraystretch{1.5}\begin{array}{cccc}
[E-(\epsilon_{0}+\xi)]/t_{S} & -t_{L}/t_{S} \\ 
1 & 0 
\end{array}
\right) \nonumber \\ 
\mbox{\boldmath $M_{\gamma,A}$} &=& \left( \def\arraystretch{1.5}\begin{array}{cccc}
[E-(\epsilon_{0}+\xi)]/t_{L} & -t_{S}/t_{L} \\ 
1 & 0 
\end{array}
\right)
\label{arma}
\end{eqnarray}
respectively.

A similar set of transfer matrices for the decoupled $B$-arm reads, 
\begin{eqnarray}
\mbox{\boldmath $M_{\alpha,B}$} &=& \left( \begin{array}{cccc}
[E-(\epsilon_{0}-\xi+\dfrac{\lambda^2}{E-\epsilon_{\mu}})]/t_{L} & -1 \\ 
1 & 0 
\end{array}
\right) \nonumber \\
\mbox{\boldmath $M_{\beta,B}$} &=& \left( \def\arraystretch{1.5}\begin{array}{cccc}
[E-(\epsilon_{0}-\xi)]/t_{S} & -t_{L}/t_{S} \\ 
1 & 0 
\end{array}
\right) \nonumber \\ 
\mbox{\boldmath $M_{\gamma,B}$} &=& \left( \def\arraystretch{1.5}\begin{array}{cccc}
[E-(\epsilon_{0}-\xi)]/t_{L} & -t_{S}/t_{L} \\ 
1 & 0 
\end{array}
\right)
\label{armb}
\end{eqnarray}
%%%%%%%%%%%%%%%%%%%%%%%%%%%%%%%%%%%%%%%%%%%%%%%
$\mbox{\boldmath $M_{\alpha}$}$ and 
$\mbox{\boldmath $M_{\gamma\beta} = M_{\gamma}.M_{\beta}$ }$ 
follow an arrangement in the Fibonacci sequence~\cite{kohmoto}.

Speaking in terms of the individual decoupled Fibonacci chains $A$ 
and $B$, we already know that each such chain offers a singular 
continuous spectrum. If we set $\lambda=0$, 
then each chain describes a separate Fibonacci lattice with 
$\epsilon_{\alpha} = \epsilon_{\beta} = \epsilon_{\gamma} = \epsilon_{0} \pm \xi$, and 
$t_{L} \ne t_{S}$. The total transfer matrix for any $m+1$-th generation of such a chain 
is related to its predecessors at the $m$th and $m-1$th generations by the matrix 
map $\mbox{\boldmath $M_{m+1} = M_{m-1}.M_{m}$}$, which will exhibit a 
{\it six cycle}~\cite{kohmoto} at $E = \epsilon_{0} \pm \xi$ for the decoupled $A$ and 
$B$ chains respectively. The existence of a six cycle of the matrix map implies that 
the above energy values are indeed eigenvalues of the infinite decoupled chains, 
and hence, the infinite ladder network as well.

Thus, the first fruit of the decoupling scheme is that, with 
$\lambda = 0$ we obtain at least two eigenvalues which 
definitely belong to the spectrum of the infinite ladder. Extending 
this idea to an $n$-strand Fibonacci ladder we can easily identify 
$n$ different energy eigenvalues of the full quasi $1$-d Fibonacci 
network in the absence of $\lambda$ (i.e. detaching the QD's) 
in a completely analytical way. The task becomes non-trivial when one 
consider the QPLN as a whole, without decoupling. Now let us look at the more
serious effect when we include the QD by making $\lambda$ non-zero.
%%%%%%%%%%%%%%
\subsubsection{The effect of QD-strand coupling ($\lambda \ne 0$)}

Let us choose Eq.~\eqref{arma}. We set $\epsilon_{\mu} = \epsilon_{0} + 
\xi$. It can easily be verified that, with this potential, the commutator 
$[\mbox{\boldmath $M_{\alpha},M_{\gamma\beta}$}] = 0$ for the 
decoupled $A$-arm, {\it irrespective of 
the energy $E$} of the electron, if $\lambda = \sqrt{t_{S}^2 - t_{L}^2}$. 
This implies that, with these conditions the individual $\alpha$-sites 
and the $\beta\gamma$ dimers in the decoupled $A$-arm can be arranged in any 
desired pattern, for example, in a perfectly periodic pattern. The spectrum 
offered by the decoupled $A$-arm Fibonacci chain will thus, under this 
choice of the QD potential $\epsilon_{\mu}$ and the tunnel hopping 
$\lambda$, will turn out to be {\it indistinguishable} from that of a 
perfectly periodic sequence of the $\alpha$-site (the `renormalized' 
$\alpha$-site, to be precise) and the $\beta\gamma$ doublet. However, this 
argument needs to go through an acid test which we discuss below. 
%%%%%%%%%%%%%%%%%%%%%%%%%%%%%%%%%55
\begin{figure}[ht]
\centering
\includegraphics[clip,width=7cm,angle=0]{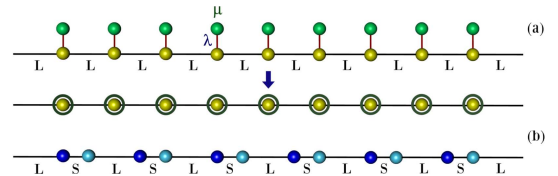}
\caption{(Color online) (a) The periodic $\alpha$+QD chain, and 
its renormalized version. (b) The $\beta\gamma$ binary ordered chain. 
The LDGF is found at the site on the horizontal line in each case. 
The QD is renormalized to produce an effective one dimensional periodic 
chain of $\bar{\alpha}$ sites, as already discussed in the text.}  
\label{chains}
\end{figure}
%*************************************************
Owing to the commutation of $\mbox{\boldmath $M_{\alpha}$}$ and 
$\mbox{\boldmath $M_{\gamma}.M_{\beta}$}$, the decoupled $A$-arm  can be thought 
to be composed of two infinite periodic chains, one with $\alpha$-sites coupled to 
the QD's, and the other, a periodic array of $\beta\gamma$ clusters. The two chains 
are shown in Fig.~\ref{chains}. Each such infinite sublattice will have its own energy 
bands. It is important to verify whether, under the special condition imposed on the 
strand-dot coupling $\lambda$, the periodic $\alpha$-lattice and the periodic 
$\beta\gamma$ lattice share the same eigenvalue spectrum. If yes, then we conclusively 
prove the existence of an absolutely continuous part in the energy spectrum of the 
original ladder network.

To check this, we first of all construct an effective one dimensional 
chain composed of renormalized $\alpha$-atoms, obtained by decimating the side 
coupled dot. The resulting chain is illustrated in Fig.~\ref{chains}(a) with encircled 
lattice points. Each such lattice point has the renormalized on-site potential equal to 
$\bar{\epsilon}_{\alpha} = \epsilon_{0}+\xi+\lambda^2/(E-\epsilon_{\mu})$. 
Each atom in the $\beta\gamma$ chain has the potential $\epsilon_{0}+\xi$.
We now evaluate the local density of states (LDOS) 
at any site of the renormalized (encircled) $\alpha$-chain ( let's call it the 
$\bar{\alpha}$ chain ) and 
any site in the periodic $\beta\gamma$ chain (Fig.~\ref{chains}(b)). 
The local densities of states are given, for these two infinite periodic lattices, and 
for a given set values of $\epsilon_0$, $\xi$, $t_{L}$ and $t_{S}$, by
$\rho_{\bar{\alpha}}(E) = (-1/\pi) [{\mathcal{F_\alpha}} (E,\lambda,\epsilon_\mu)]^{-1/2}$ and 
$\rho_{\beta\gamma} = (-1/\pi) [{\mathcal{G_{\beta\gamma}}}(E)]^{-1/2}$, where,
%%%%%%%%%%%%%%%%%%%%%%%
\begin{equation}
\mathcal{F_{\alpha}}(E,\lambda,\epsilon_{\mu}) = 4 t_{L}^2 - ( E - \epsilon_{0} - \xi - 
\frac{\lambda^2}{E - \epsilon_{\mu}} )^2 
\end{equation}
and,
\begin{equation}
\mathcal{G_{\beta\gamma}}(E) = \frac{4 t_{L}^2 t_{S}^2}{(E - \epsilon_{0} - \xi)^2} - 
( E - \epsilon_{0} - \xi - \frac{t_{L}^2 + t_{S}^2}{E - \epsilon_{0} - \xi} )^2 \\
\end{equation}

If we now set 
$\epsilon_{\mu} = \epsilon_{0} + \xi$, then a simple algebra yields that, the difference 
$\Delta = \mathcal{G_{\beta\gamma}} - \mathcal{F_{\alpha}}$ is given by,
\begin{equation}
\Delta = \frac{\left[2 ( E-\epsilon_{0} - \xi )^2 + t_{L}^2 - t_{S}^2 
 - \lambda^2 \right] ( \lambda^2 + t_{L}^2 - t_{S}^2 )}{( E - \epsilon_{0} - \xi )^2}
\label{den}
\end{equation}
which clearly vanishes for any value of the energy of the electron 
the moment we tune $\lambda = \sqrt{t_{S}^2 - t_{L}^2}$.
Thus the LDOS's of the two basic constituent periodic chains 
become identical {\it independent of energy}, resulting in a complete overlap of the 
spectra of these individual chains under the above resonance condition. 
As each of the $\alpha$ and the $\beta\gamma$ sublattices provide a two subband 
continuum in their energy spectra (an obvious effect of the binary periodic lattice), 
it is obvious that these two spectral continua will retain their positions in the 
full spectrum of the Fibonacci QPLN when the QD potential and the QD-strand tunnel 
coupling are tuned to the above special values.

It should be appreciated that once we fix $\epsilon_{\mu} = 
\epsilon_{0} + \xi$, the set of Eq.~\eqref{armb} corresponding to 
the decoupled $B$ strand, represents a mixed model linear Fibonacci chain. 
Its contribution to the full energy spectrum is the usual fragmented, 
multi-fractal one. As the full spectrum of the two-arm ladder will be 
a convolution of both, we expect it to have absolutely continuous parts 
decorated with a spiky envelope, and a mixture of {\it extended} 
Bloch-like states and the {\it critical} eigenfunctions which are 
characteristic of a quasi-periodic Fibonacci sequence of potentials.
This is precisely what we observe in Fig.~\ref{2stranddos}(a).

At this point one must be careful to interpret  Fig.~\ref{2stranddos}(a).
In the full energy spectrum, critical and extended eigenstates can never 
coexist. So, even in the full spectrum of the un-decoupled QPLN 
the entire range shown in Fig.~\ref{2stranddos}(a) by grey 
shade corresponds to extended Bloch-like states only. Beyond it, every 
wave function will be critical in nature. The contributions coming from 
each decoupled strand are only shown separately in this figure.

\subsubsection{The RSRG analysis}

The above arguments are confirmed by an exact real space renormalization 
group (RSRG) calculation of the local density of states, first of each 
individual decoupled strands and then of the full two-arm Fibonacci ladder 
without decoupling it. The first task is already discussed in literature~\cite{ashraff,arunava}, 
and need not be talked about again in here. We instead, provide the 
details of the RSRG scheme of the full ladder network. We 
renormalize the full two-strand ladder by decimating the 
$\beta$-vertices. This generates {\it new} hopping integrals across 
the diagonals of the fundamental rectangular plaquettes of 
renormalized ladder denoted by $d_{L}^{\prime}$ and $d_{S}^{\prime}$ in the 
equations to follow. To implement the decimation we need to assign 
three different symbols to the inter-arm hopping $\xi$, viz., $\xi_{\alpha}$, 
$\xi_{\beta}$ and $\xi_{\gamma}$ connecting the $\alpha$, $\beta$ and 
$\gamma$ vertices of the two strands. Of course, we begin with 
$\xi_{\alpha}=\xi_{\beta}=\xi_{\gamma}=\xi$, and $d_{L}=d_{S}=0$. The recursion relations 
arising out of this decimation procedure are given by, 
%%%%%%%%%%%%% INSERT RECURSION RELATIONS HERE%%%%%%%%%%%
\begin{eqnarray}
{\bar{\epsilon}}_{\alpha}^{\prime}&=&\epsilon_{\gamma}+\mathcal{A}t_{L}+\mathcal{B}t_{S}
+\mathcal{C}d_{L}+\mathcal{D}d_{S}\nonumber \\
\epsilon_{\beta}^{\prime}&=&\epsilon_{\gamma}+\mathcal{B}t_{S}+\mathcal{D}d_{S}
\nonumber \\
\epsilon_{\gamma}^{\prime}&=&\bar{\epsilon}_{\alpha}+\mathcal{A}t_{L}+
\mathcal{C}d_{L}\nonumber \\
t_{L}^{\prime}&=&\mathcal{B}t_{L}+\mathcal{D}d_{L}\nonumber \\
t_{S}^{\prime}&=&t_{L}\nonumber \\
d_{L}^{\prime}&=&\mathcal{B}d_{L}+\mathcal{D}t_{L}\nonumber \\
d_{S}^{\prime}&=&d_{L}\nonumber \\
\xi_{\alpha}^{\prime}&=&\xi_{\gamma}+\mathcal{A}d_{L}+\mathcal{B}d_{S}
+\mathcal{C}t_{L}+\mathcal{D}t_{S}\nonumber \\
\xi_{\beta}^{\prime}&=&\xi_{\gamma}+\mathcal{B}d_{S}+\mathcal{D}t_{S}
\nonumber \\
\xi_{\gamma}^{\prime}&=&\xi_{\alpha}+\mathcal{A}d_{L}+
\mathcal{C}t_{L} 
\label{recur}
\end{eqnarray}
where
\begin{eqnarray}
\mathcal{A}&=&\dfrac{(E-\epsilon_{\beta})t_{L}+\xi_{\beta}d_{L}}
{\mathcal{W}}\nonumber \\
\mathcal{B}&=&\dfrac{(E-\epsilon_{\beta})t_{S}+\xi_{\beta}d_{S}}
{\mathcal{W}}\nonumber \\
\mathcal{C}&=&\dfrac{(E-\epsilon_{\beta})d_{L}+\xi_{\beta}t_{L}}
{\mathcal{W}}\nonumber \\
\mathcal{D}&=&\dfrac{(E-\epsilon_{\beta})d_{S}+\xi_{\beta}t_{S}}
{\mathcal{W}}\nonumber \\
\mathcal{W}&=&(E-\epsilon_{\beta})^{2}-\xi_{\beta}^{2}\nonumber \\
\bar{\epsilon}_{\alpha} &=& \epsilon_{0} + 
\lambda^2/(E -\epsilon_{\mu}) \nonumber
\end{eqnarray}
%%%%%%%%%%%%%%%%%%%%%%%%%%%%%%%%%%%%%%%%%%%%%%%%%%%%%%%%

A small imaginary part is added to the energy $E$, and the LDOS at 
an $\alpha$, $\beta$ or $\gamma$ site is obtained by calculating the 
respective local Green's function $G_{00}^{i}=(E-\epsilon_{i}^*)^{-1}$, 
where $i=\alpha, \beta$ or $\gamma$. $\epsilon_{i}^*$ represents the 
fixed point of the respective on-site potential as the hopping integrals 
flow to zero under the RSRG iterations. The LDOS is given by, 
\begin{equation}
\rho_{i} = -\dfrac{1}{\pi}\text{Im}[G_{00}^i] 
\end{equation}

\begin{figure}[ht]
\centering
\includegraphics[clip,width=7cm,angle=0]{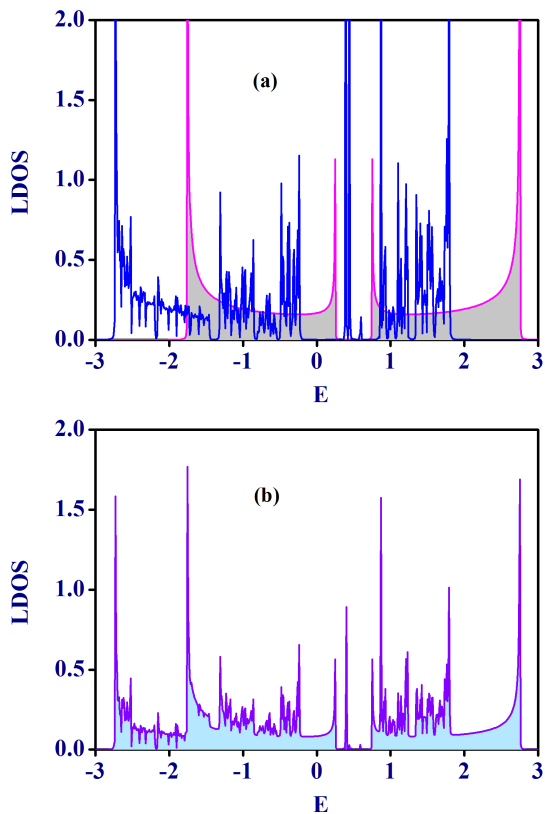}
\caption{(Color online) Plot of local density of states 
(LDOS) at an $\alpha$-vertex for a two-strand ladder network. 
The upper panel shows the LDOS spectra of the 
decoupled individual linear chains, one chain satisfying 
resonance condition. The lower panel shows the LDOS spectrum 
of the composite ladder network as obtained from the 
RSRG scheme, under resonance condition. 
The values of the parameters chosen are 
$\epsilon_{\alpha}=\epsilon_{\beta}=\epsilon_{\gamma}=0$,
${\epsilon_{\mu}}=0.5$, $t_{L}=1$, $t_{S}=1.25$, $\lambda=0.75$, and $\xi=0.5$, 
measured in unit of $t_{L}$.}
\label{2stranddos}
\end{figure}

The results are presented in Fig.~\ref{2stranddos}. In panel (a) the 
spectra of the decoupled $A$- and $B$-arms are shown separately, 
while panel (b) illustrates the spectrum of the full two-arm ladder as obtained 
from the RSRG study. The existence of the absolutely continuous portions in the 
energy spectrum is obvious. The character of the eigenfunctions belonging to the 
spectrum is checked from an observation of the flow of the the hopping 
integrals $t_{L}$, $t_{S}$ and the diagonal hopping amplitude under RSRG 
iterations. For any energy eigenvalue, chosen from any portion of 
the absolutely continuous sub-bands (grey shaded portion in 
Fig.~\ref{2stranddos}(a)) all the hopping integrals, as 
obtained from Eq.~\eqref{recur}, remain non-zero 
for an arbitrarily large number of iterations confirming the extended 
character of the wave functions. Beyond the extended regime, the hopping 
integrals flow to zero under iteration indicating that the eigenfunctions 
are not extended. In Fig.~\ref{2stranddos}(b) we see the full LDOS spectrum 
from the coupled two-strand QPLN 
obtained from the RSRG method
(using the set of 
Eq.~\eqref{recur}). 
The light blue shade in the spectrum is just to highlight the 
region of non-zero LDOS. The extended eigenstates occupy the precise 
positions as depicted by the grey shaded part in Fig.~\ref{2stranddos}(a).
The eigenstates form a mixed spectrum of 
critical and extended states, as one traverses the energy axis from the left. 
%*************************************************
%*************************************************

An additional support in regard of the continuous subbands and the 
extended eigenstates is also obtained from an observation of the flow 
pattern of the entire parameter space of the decoupled strands that 
satisfies the condition $\epsilon_{\mu}=\epsilon_{0}+\xi$ and 
$\lambda=\sqrt{t_{S}^{2}-t_{L}^{2}}$. For such a choice, under 
successive RSRG step, we always get 
$\bar{\epsilon}_{\alpha}^{(j)} \neq 
\epsilon_{\beta}^{(j)}=\epsilon_{\gamma}^{(j)}$ and 
$t_{L}^{(j)} \neq t_{S}^{(j)}$ for the decoupled $A$-strand. 
This is a typical flow in the parameter space 
that corresponds to the extended eigenstates in quasiperiodic Fibonacci 
chains~\cite{arunava2}. In the present case it confirms the extended Bloch-like 
character of the wave function for the entire energy regime (given by 
the grey color in Fig.~\ref{2stranddos}(a)).
\subsection{Three-arm Fibonacci ladder and beyond}

The motivation behind extending the above analysis to a three-arm Fibonacci ladder 
stems
%*************************************************
\begin{figure}[ht]
\centering
\includegraphics[clip,width=8cm,angle=0]{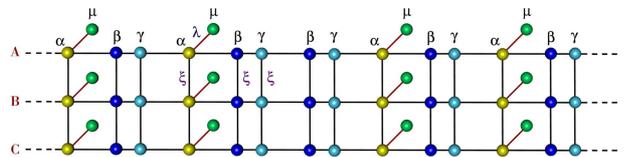}
\caption{(Color online) Schematic diagram of a three-strand Fibonacci ladder network.
Meaning of all other symbols are same as in figure~\ref{lattice1}.}
\label{lattice2}
\end{figure}
%*************************************************
from the curiosity to know whether, as one extends the ladder network in the 
transverse direction, any possibility of a re-entrant transition (or, crossover) 
from a singular continuous to an absolutely continuous energy spectrum becomes 
apparent. For disordered ladders comprising the same $\alpha$ and $\beta\gamma$ 
clusters this would imply the possibility of an {\it insulator-metal} transition 
that can be triggered by an appropriate choice of the numerical values of the QD 
potential $\epsilon_{\mu}$ and the tunnel hopping integral $\lambda$. 

To check such a possibility we decouple a three-arm Fibonacci ladder 
(Fig.~\ref{lattice2}). We follow the same prescription for decoupling 
as outlined in the case of a two-strand ladder. Diagonalization 
of the potential matrix $\mbox{\boldmath $\epsilon_{i}$}$ now 
yields three eigenvalues, viz., $\epsilon_{0}$ and 
$\epsilon_{0} \pm \sqrt{2}\xi$, and the effective potential at the 
$\alpha$-site in each decoupled arm will be 
$\epsilon_{0}+\lambda^{2}/(E-\epsilon_{\mu})$ (for arm $A$, say), 
$\epsilon_{0}+\sqrt{2}\xi+\lambda^{2}/(E-\epsilon_{\mu})$ (for arm $B$, say) 
and $\epsilon_{0}-\sqrt{2}\xi+\lambda^{2}/(E-\epsilon_{\mu})$ (for arm $C$). 
We now end up having 
three sets of equations, one for each decoupled strand. For example, 
\begin{eqnarray} 
&\left [E - \left(\epsilon_{0}+\dfrac{\lambda^2}{E-\epsilon_{\mu}}\right)\right ] 
\phi_{i,A} = t_{L} \phi_{i+1,A} + t_{L} \phi_{i-1,A} \nonumber \\
&(E - \epsilon_{0}) \phi_{i,A} = t_{S} \phi_{i+1,A} + t_{L} \phi_{i-1,A} \nonumber \\
&(E - \epsilon_{0}) \phi_{i,A} = t_{L} \phi_{i+1,A} + 
t_{S} \phi_{i-1,A} 
\label{3decouple1}
\end{eqnarray}
corresponding to the $\alpha$, $\beta$ and $\gamma$ sites respectively 
in the decoupled $A$-strand. Two sets of similar equations are also obtained 
for the decoupled $B$- and $C$-strands, with the effective potential 
at the $\alpha$-site being equal to 
$\epsilon_{0}+\sqrt{2}\xi+\lambda^2/(E-\epsilon_{\mu})$ and 
$\epsilon_{0}-\sqrt{2}\xi+\lambda^2/(E-\epsilon_{\mu})$ respectively. 
The effective potentials at the $\beta$ and $\gamma$ sites in the  
$B$ and $C$ strand are $\epsilon_{0} \pm \sqrt{2}\xi$ as there is 
no side coupling to these sites.

\begin{figure}[ht]
\centering
\includegraphics[clip,width=7cm,angle=0]{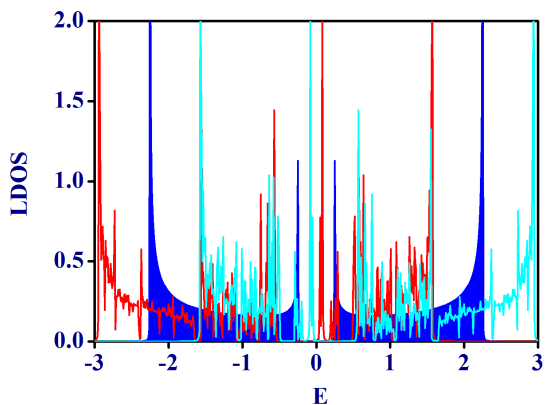}
\caption{(Color online) Plot of LDOS for a three-strand 
Fibonacci ladder network. It represents the 
LDOS spectra of the three decoupled isolated chains, one satisfying 
resonance condition (in blue shaded color) and the other two with 
off-resonance condition (in red and cyan color). The parameters 
chosen here are same as in Figure~\ref{2stranddos}.}
\label{3stranddos}
\end{figure}

If, as before, we set $\epsilon_{\mu}=\epsilon_{0}$ 
in Eq.~\eqref{3decouple1}, then 
$[\mbox{\boldmath $M_{\alpha},M_{\gamma\beta}$}]=0$ for the decoupled 
$A$-strand. The contribution to the full spectrum coming from this 
separate chain will be two absolutely continuous sub-bands. With the 
potential of the QD set at $\epsilon_{\mu}=\epsilon_{0}$, the two 
other decoupled strands represent two different Fibonacci chains 
in the mixed model. The spectrum resulting from each of them will 
remain fragmented Cantor type. The overall spectrum, as we now have 
known, will be a convolution of the three.
The absolutely continuous portions will still hold their positions 
in the main spectrum, and will be flanked by the spiky, fragmented parts 
characteristics of critical eigenstates of a Fibonacci lattice 
(Fig~\ref{3stranddos}). The eigenfunctions belonging to the continuous 
sub-bands will, needless to say, be of extended character.
%*************************************************
%************************************************* 

One needs to appreciate that each spectrum offered separately by the 
decoupled $B$- and $C$-strands will have different ``band-centres" around 
which the singular continuous spectra of the individual decoupled 
Fibonacci chain will spread out. With an $n$-strand ladder model, 
there will be $n-1$ such centers from the $n-1$ 
decoupled strands around which Cantor set energy eigenvalues  will be 
distributed, the remaining strand contributing the absolutely continuous 
sub-bands. The centres will be densely populated around the outer 
edge of the continuous sub-bands as $n \to \infty$. Thus, with 
increasing number of strands there is a possibility of observing a 
smooth critical to extended state crossover in the spectrum of such 
quasi $1$-d aperiodic ladder models. Clearly, the argument prevails 
even if we have an $n$-strand off diagonally disordered ladder network. 
The side coupled QD's with appropriately tuned potential one can then 
expect a re-entrant {\it insulator-metal} transition.

Before leaving this discussion, it is pertinent to mention that, the 
commutaion of transfer matrices leading to a complete, energy independent 
delocalization of electronic states have been reported recently in certain 
special examples of quasi-one dimensional lattices~\cite{pal1}. But, those 
systems do not open up any possibility of an insulator-metal transition, 
in the current spirit of the work. At the same time, it is useful to note 
that, the commutivity of transfer matrices has been used in the past to 
unravel discrete unscattered eigenstates in an array of quantum wells~\cite{adame}
 or a disordered distribution of P\"{o}schl-Teller potential~\cite{alberto2}. The 
difference with the present communication and these two latter references lies 
in the energy-independent character of the matrix-commutation that plays the 
key role.
%%%%%%%%%%%%%%%%%%%%%%%%%%%%%%%%%%%%%%%%%%%%%%%%%%%%%%%%%%%%%%%%%%%%%%%%%%%%%%%%%%
\section{Two terminal transmission}
\label{sec3}
\begin{figure}[ht]
\centering
\includegraphics[clip,width=8.5cm,angle=0]{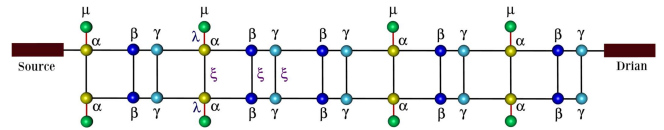}
\caption{(Color online) Schematic view of a finite size two-strand QPLN 
clamped between the source and the drain.}
\label{lattice3}
\end{figure}

We now present partial results (to save space) 
for the two terminal 
transmission coefficient of a finite two strand ladder without 
decoupling it. No change of basis is made, and the full 
finite QPLN is treated within the RSRG scheme.
The basic working formula is already elaborated in 
the existing literature~\cite{stone}. The ladder is clamped between two 
semi-infinite perfectly ordered leads (viz., the source and the drain) 
at any two extremities (Fig.~\ref{lattice3}). 
%*************************************************
%*************************************************
The leads are characterized by uniform on-site potential $\epsilon_{\mathcal{L}}=0$ and 
nearest neighbor hopping integral $\tau_{\mathcal{L}}=2$.
The portion 
of the ladder clamped between the leads is renormalized to bring it down to a 
{\it diatomic molecule}, and the transmission coefficient 
is easily obtained~\cite{arunava3,pal2}. The procedure is standard and to save space again, we do 
not provide the details. 

It should be appreciated that, the precise details of the 
transmission coefficient definitely depends on which points in the 
ladder the leads are attached to.
Only one result 
is shown here in which the two semi-infinite ordered leads are connected to the 
extreme points of the upper strand of a $10$-th generation Fibonacci ladder 
having $89$ bonds in each arm.
%*************************************************
\begin{figure}[ht]
\centering
\includegraphics[clip,width=7cm,angle=0]{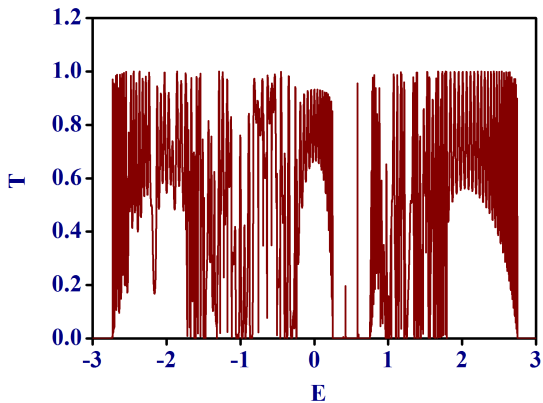}
\caption{(Color online) Transmission characteristics of 
a $10$-th generation two-strand Fibonacci ladder network under 
resonance condition. Both the incoming and outgoing leads, 
viz., the source and the drain are connected in the upper strand. 
System parameters chosen are 
$\epsilon_{\alpha}=\epsilon_{\beta}=\epsilon_{\gamma}=0$, 
${\epsilon_{\mu}}=0.5$, $t_{L}=1$, $t_{S}=1.25$, $\lambda=0.75$, and $\xi=0.5$, and 
the lead parameters are $\epsilon_{\mathcal{L}}=0$ and $\tau_{\mathcal{L}}=2$.}
\label{transport}
\end{figure}
%*************************************************
One needs to appreciate that, even in the absence of any attached 
QD, a two strand finite QPLN can exhibit high transmission spikes with 
$T$ close to unity. This normally results from the finite size of the system, as 
well as from the fact that a ladder network consists of loops that brings in 
a flavor of long range connection, and may well be responsible for quantum 
interference leading to high transmittivity at selected values of electron-energy.
Also, the attachment to the leads, and the difference between the hopping 
integral in the lead and that in the bulk of the system give rise to strong scattering 
effects at the contacts, affecting the overall transmission of the QPLN system.
Therefore, the results obtained from a finite size analysis should be 
carefully compared with the predictions which come from the anatomy of an 
infinite system.
 
Nevertheless, the existence of the continuous parts in the energy spectrum of the 
infinite system is already showing up in the enhanced value of the 
transmission coefficient at appropriate values of the electron 
energy as can be seen in Fig.~\ref{transport}. 
The apparent drop to close to zero value at certain regions even within the
continuous zone is due to the scale at which the transmission coefficient 
is shown here. However, the transmission coefficient exhibits wide fluctuations 
owing to the quasiperiodic nature of network. The effect of quantum interference 
is manifestly observed here. 
The finite value of the transmission coefficient 
even beyond the absolutely continuous part of the spectrum is due to the finite 
size of the system, and gets lower and lower, and 
the spectrum gets more spiky with increasing system size. Also, one can anticipate 
eigenstates with their 	`localization length' much larger compared to the 
size of the system as considered in the present calculation.
%%%%%%%%%%%%%%%%%%%%%%%%%%%%%%%%%%%%%%%%%%%%%%%%%%%%%%%%%%%%%%%%%%%%%%%%%%%%%%%%%%
\section{Concluding remarks}
\label{sec4}
We have shown that Fibonacci quasi-periodic ladder networks, where 
quasi-periodic order is introduced in the distribution of bonds, 
keeping the potential at every vertex same, are capable of 
producing absolutely continuous portions in their energy spectra. 
This requires the attachment of quantum dots from one side to a 
special set of sites in each arm of the ladder. A suitable adjustment 
of the dot potential, together with the dot-backbone coupling can 
generate absolutely continuous energy sub-bands, decorated at the 
flanks by the usual critical states -- a characteristic of 
quasi-periodic lattices. Apart from opening up the possibility 
of engineering the transport of excitons in such multi-strand 
systems, a very basic issue of Anderson localization is 
reviewed. It seems that, the existence of localized eigenstates in 
similar disordered ladder networks may strongly depend on the 
numerical values of the parameters in the Hamiltonian.

The analysis is carried over to other quasi-periodically ordered 
systems, such as the copper mean chain, the aperiodic Thue Morse chain and even to 
the disordered ladder networks with locally and non-locally coupled 
dots. The conditions for the existence of a continuum in the energy spectrum becomes 
much more non-trivial in such systems. These issues will be reported in a forthcoming 
article.   
%%%%%%%%%%%%%%%%%%%%%%%%%%%%%%%%%%%%%%%%%%%%%%%%%%%
\begin{acknowledgments}
B.P. is grateful to DST, India for an INSPIRE Fellowship. A.C. acknowledges 
financial support from DST, India through a PURSE grant, and The Abdus Salam 
International Centre for Theoretical Physics, Trieste, Italy for its hospitality 
and financial support.
\end{acknowledgments} 
%%%%%%%%%%%%%%%%%%%%%%%%%%%%%%%%%%%%%%%%%%%%%%%%%%%

%%%%%%%%%%%%%%%%%%%%%%%%%%%%%%%%%%%%%%%%%%%%%%%%%%%
\end{document}